\documentclass{article}

\usepackage{icrctc07}       

\newcommand{\eV}{\,\mathrm{eV}}

\newcommand{\GeV}{\,\mathrm{GeV}}
\newcommand{\muG}{\,\mu\mathrm{G}}
\newcommand{\kpc}{\,\mathrm{kpc}}

\newcommand{\cm}{\,\mathrm{cm}}

\newcommand{\g}{\,\mathrm{g}}

\newcommand{\yr}{\,\mathrm{yr}}

\begin{document}

\title{Numerical Propagation of Cosmic Rays in the Galaxy}
\authors{Daniel De Marco$^1$, Pasquale Blasi$^2$, Todor Stanev$^1$}
\shorttitle{Numerical Propagation of Cosmic Rays in the Galaxy}
\shortauthors{D. De Marco et al.}

\afiliations{%
$^1$ Bartol Research Institute, University of Delaware, Newark, DE 19716, USA\\
$^2$ INAF/Osservatorio Astrofisico di Arcetri, Largo E. Fermi 5, 50125 Firenze, ITALY%
}

\email{ddm@bartol.udel.edu}

\abstract{%
We present a Monte-Carlo (MC) calculation of the propagation of cosmic ray protons in the Galaxy for
energies above 1 PeV. We discuss the relative strengths of competing effects such as
parallel/perpendicular diffusion and drifts in toy models of the Galaxy. We compare our estimates
with the results of the MC calculation for the toy models and then we apply the MC calculation to a
few more realistic models of the Galactic magnetic field. We study the containment times in
different models of the magnetic field in order to understand which one may be consistent with the
low energy data.}

\maketitle

\section{Introduction}
The transport of cosmic rays in the Galaxy is usually studied solving the diffusion-convection
equation from a distribution of sources in a medium with given diffusion properties. This approach,
although very successful (see Ref.~\cite{moska} for a recent review), can not be applied at
arbitrarily high energy because at some point the diffusion approximation breaks down. In the
Galaxy this happens, for protons, around $10^{17}\eV$. The study of CR transport in this energy
region is performed using numerical simulations of the propagation of single particles in the
Galactic magnetic field (GMF). Besides the obvious advantage of being able to study the transition
region, another advantage of the latter method is the ability to use more realistic models of the
GMF, including the arms, all the gradients and so on. The big drawback of this approach is in the
computing time that makes it usable only at high energy ($10^{15}\eV$ and above).

From observations at low energy, $\sim\GeV$, the residence time of particles in the Galaxy is
estimated at about $10^7\yr$ with an energy dependence of $E^{-0.6}$. When extrapolated to higher
energies, this trend produces nonsensical results, predicting, for example, huge anisotropies
already around $10^{15}\eV$ where they are not observed. Assuming an energy dependence of
$E^{-1/3}$, as one would expect from Kolmogorov turbulence, the extrapolations are much better, but
still a few times larger than the experimental results \cite{hillasrev}. 

In principle one would want to use the diffusion-convection equation at low energy, the numerical
simulations at high energy and possibly match the two results in between. Up to now, the simulations
at high energy \cite{hisimu} were not successful in this respect. The containment time at
$10^{17}\eV$ is calculated as $10^5\yr$ with an energy dependence of $1/E$. Extrapolated to lower
energy, this result produces escape times exceeding the age of the universe.

In order to investigate this discrepancy, we developed a numerical simulation of the propagation of
particles in arbitrary magnetic fields and we applied it to several toy models of the Galaxy to
separately study the various effects contributing to the transport, such as parallel and
perpendicular diffusion, drifts and so on. For details on the simulation code see Ref.~\cite{DMBS07}.

\section{Diffusion Coefficients}
We calculate the diffusion coefficients in a magnetic field composed of a constant regular field in
the $z$ direction and a turbulent field with a Kolmogorov spectrum. We inject particles in this
field and we follow their trajectories recording their positions as a function of time. Fitting the
distribution of the particle positions as a function of time we calculate the diffusion coefficients
along the three axes. For the parallel diffusion coefficient we find the usual energy dependence of
$E^{1/3}$ at low energy and $E^2$ at high energy, whereas for the perpendicular one we find a
steeper dependence at low energy: $E^{0.5-0.6}$. This is already an interesting result because it
means that in particular geometries where the transport is dominated by perpendicular diffusion one
can obtain, from a Kolmogorov spectrum of turbulence, a diffusion coefficient, and therefore a
residence time, steeper than the usual one.

\section{Toy Models}
\begin{figure}
  \centering
  \includegraphics[width=0.45\textwidth]{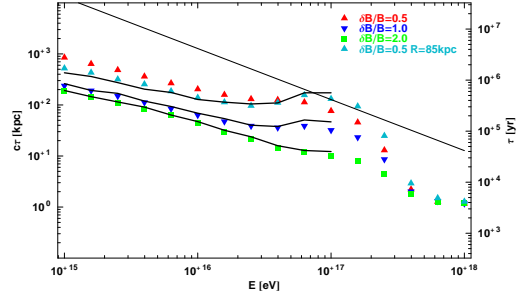}
  \caption{Times of escape in the azimuthal field. The colored points are the results of the
  simulations for different scenarios. The thin lines are expected escape times calculated with the
  diffusion coefficient, whereas the thick straight line is the timescale for
  drifts.
  }\label{fig:toe1}
\end{figure}

The first toy model we consider has a purely azimuthal magnetic field that is constant, $1\muG$,
everywhere. Superimposed to this regular field there is a turbulent component whose magnitude is
proportional to the one of the regular field with a proportionality constant of $0.5$, $1$ or $2$.
We inject particles at the solar system position ($8.5\kpc$ from the center) and we record the time
required for escaping a cylinder with half-height of $0.5\kpc$ and radius of $10\kpc$. It is important
to note that in this scenario the transport is dominated by the perpendicular diffusion and possibly
by drifts, and that the parallel diffusion can be completely neglected since it scatters the
particles back and forth along the field lines, but it does not help them escaping the disk since the
field lines are closed. The times of escape are plotted with colored points in
Fig.~\ref{fig:toe1}. The red upward, blue downward triangles and green squares correspond to three
different levels of turbulence: $\delta B/B=0.5,1,2$ respectively\footnote{Please note that here and
in the following when denoting $\delta B$ we actually mean: $\sqrt{\langle\delta
B^2\rangle}$.}. The thick solid straight line represents the drift time-scale calculated for the
average drift velocity, whereas the thin solid lines are the time-scales for perpendicular diffusion;
they are proportional to $h^2/D_\perp$. The perpendicular diffusion coefficient is the one
calculated as described in the previous section. As it is clear from Fig.~\ref{fig:toe1} the
transport description as perpendicular diffusion is pretty good for the two cases $\delta B/B=1$ and
$2$, but it is not very accurate for $\delta B/B=0.5$. The discrepancy is at low energy, around
$10^{15}\eV$, and in this region the drift time-scale is much bigger than the diffusion one and then
this effect is not caused simply by drifts: they are relevant only at higher energies. The effect is
however produced by the curvature of the field lines since injecting the particles at $85\kpc$,
where the curvature is ten times smaller, produces a time of escape that is in much better accord
with the diffusion time-scale (see light blue upward triangles in Fig.~\ref{fig:toe1}).

\begin{figure}[t]
  \centering
  \includegraphics[width=0.45\textwidth]{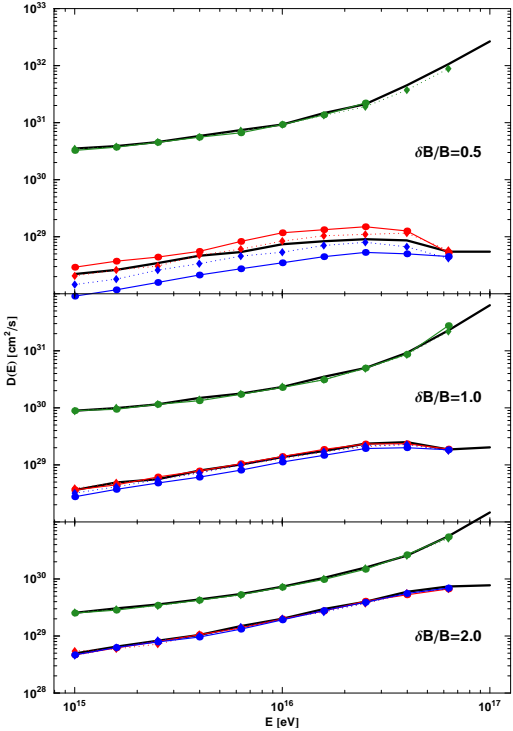}
  \caption{Diffusion coefficients in a curved regular field for the level of turbulence indicated in
  the plots. In each plot the black lines are the parallel (upper) and perpendicular (lower)
  diffusion coefficients obtained in the constant field. The green, red and blue lines are diffusion
  coefficients respectively in the azimuthal, radial and $z$ direction. Solid lines are for
  injection at $8.5\kpc$, dotted lines are for injection at $85\kpc$.}\label{fig:azdiffu}
\end{figure}

In order to better understand this effect we calculate the diffusion coefficients in the curved
magnetic field to check if the curvature of the field lines has an effect also on the
diffusion coefficients. Our results are plotted in Fig.~\ref{fig:azdiffu}. We find that with
smaller levels of turbulence the two perpendicular diffusion coefficients, in the radial and in the
$z$ direction, are modified. The first one is increased and the second one is decreased. This is
consistent with the results of Fig.~\ref{fig:toe1}, since a smaller diffusion coefficient in the $z$
direction produces a larger time of escape. Reducing the curvature of the field lines, i.e. calculating
the diffusion coefficients at larger distances from the center, the effect is reduced (see dotted
lines in Fig.~\ref{fig:azdiffu}). The effect promptly disappears increasing the level of turbulence.
Fig.~\ref{fig:azdiffu} is then telling us that the curvature of the field lines not only produces a
drift, or the rigid displacement of the distribution of the particle positions, but it also affects
the way the distribution evolves with time along the three directions.

Though keeping in mind that this is only a toy model of the magnetic field of the Galaxy, it is
interesting to note that at energy $10^{15}\eV$ the escape time is $\tau_{15}\approx 0.5-5$ million
years (the halo height here is only $0.5\kpc$). These numbers are of the same order of magnitude of
the confinement times estimated from the abundance of light element in the GeV region, which means
that in order to fit these observations one should postulate that the escape time below $10^{15}\eV$
should be practically energy independent. We could not envision any realistic mechanism able to
justify such an expectation. It follows that within the limitations of the present toy model it is very
hard to obtain a realistic, even qualitative, description of what is observed in the Galaxy at much
lower energies. This conclusion is confirmed also by the curves on the grammage that show that
at $10^{15}\eV$ cosmic rays traverse already a column density of $1-2\g\cm^{-2}$.

We modified the above toy model introducing several complications mimicking the structure of the
Galaxy, for example a gradient in radial direction or in the $z$ direction, but the general result
did not change much. The slopes of the times of escape remained quite steep, although the absolute
values became smaller due to the increase of the drifts.

\section{``Realistic'' Galactic Magnetic Fields}

The knowledge of the magnetic field in our Galaxy is very poor, both for what concerns its regular
component and more so for what concerns the turbulent one. Having this in mind we considered a very
simple model with the intent not of being very realistic, but just as an example of how the simple
picture presented in the previous section becomes suddenly much more complicated. Indeed, following
Ref.~\cite{han,stanevmf}, we considered the BSS model: it includes the spiral arms, has a radial
gradient and a gradient in the $z$ direction, but it does not have an halo field, a bar in the
center or other complications.

\begin{figure}
  \centering
  \includegraphics[width=0.45\textwidth,bb=108 94 588 416,clip]{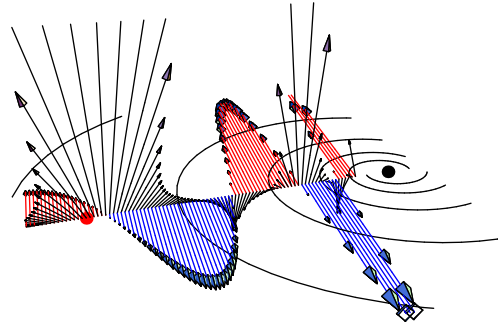}
  \caption{Galactic Magnetic Field and drifts in the BSS models. The black dot is the galactic
  center, the red one is the solar system position. The black lines show the position of the center
  of the arms. The blue and red arrows represent the magnitude and direction of the GMF in the given
  position, while the black arrows show the drift speed.}\label{fig:galdrifts}
\end{figure}

We plot the drift velocity calculated in several points along the line connecting the galactic
center (black dot) with the solar system position (red dot) in Fig.~\ref{fig:galdrifts}. The black
lines show the position of the center of the arms, the blue and red arrows represent the magnitude
and direction of the GMF in the given position, while the black arrows show the drift speed. In the
toy model presented in the previous section the drift velocity was always in the $z$ direction
and had a position dependent magnitude. In this case the situation is much more complicated with
the drift pushing the particles sometimes toward the center of the arms and sometimes toward the
inter-arms space. Moreover in this case both the magnitude and direction of the drift depend on the
position and on the pitch angle of the particle. Indeed, in the plot of Fig.~\ref{fig:galdrifts} the
drift velocity is calculated for a particle with an injection cosine with respect to the local
magnetic field of $\sqrt{2/3}$ and for a position just above the galactic plane. Calculating it for
another pitch angle or another position would produce completely different results. For example,
considering the same pitch angle, but a position just below the plane, we would obtain the opposite
sign for the radial component of the drift velocity and thus a very different picture.

\begin{figure}
  \centering
  \includegraphics[width=0.45\textwidth]{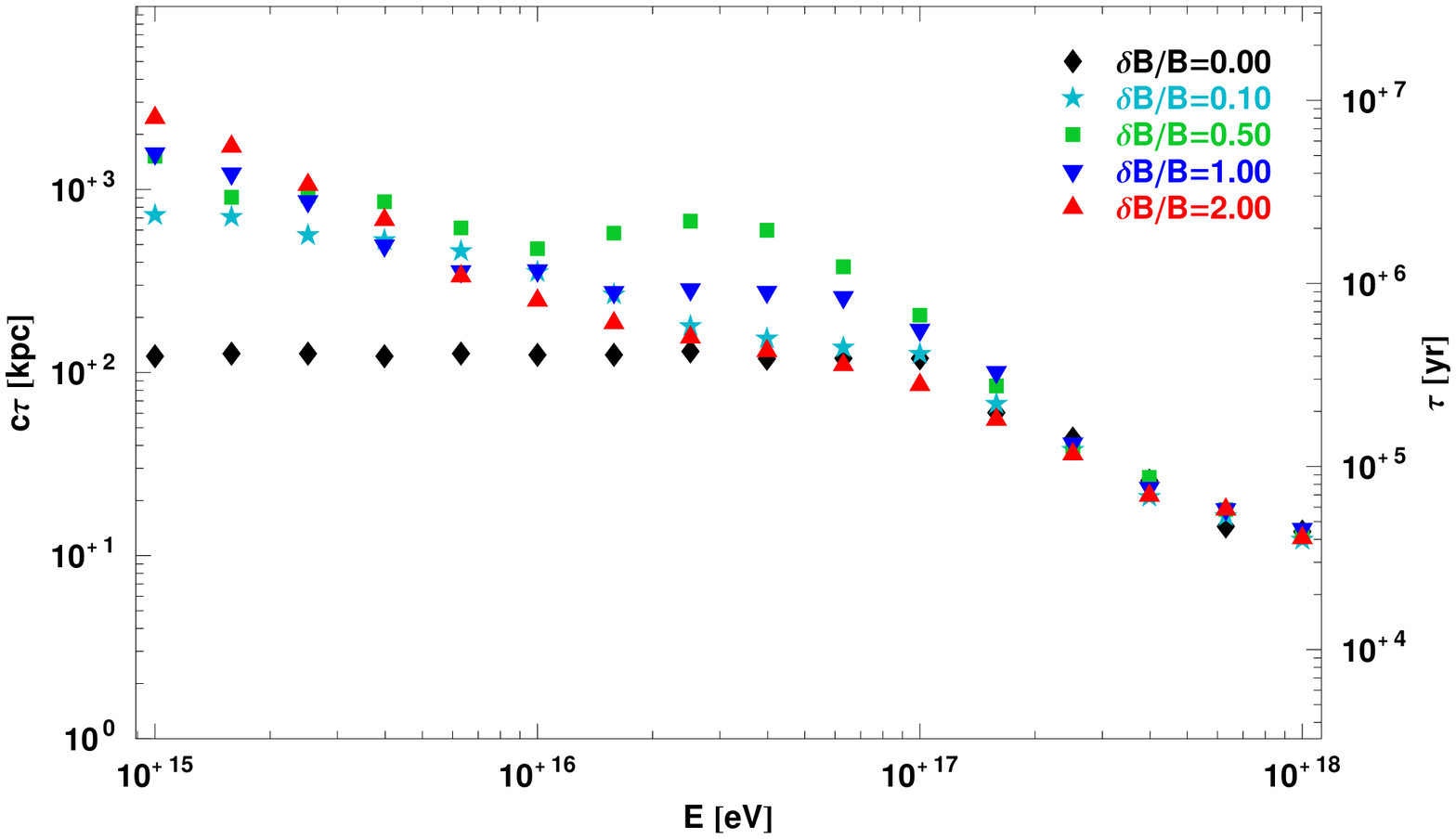}
  \quad
  \includegraphics[width=0.45\textwidth]{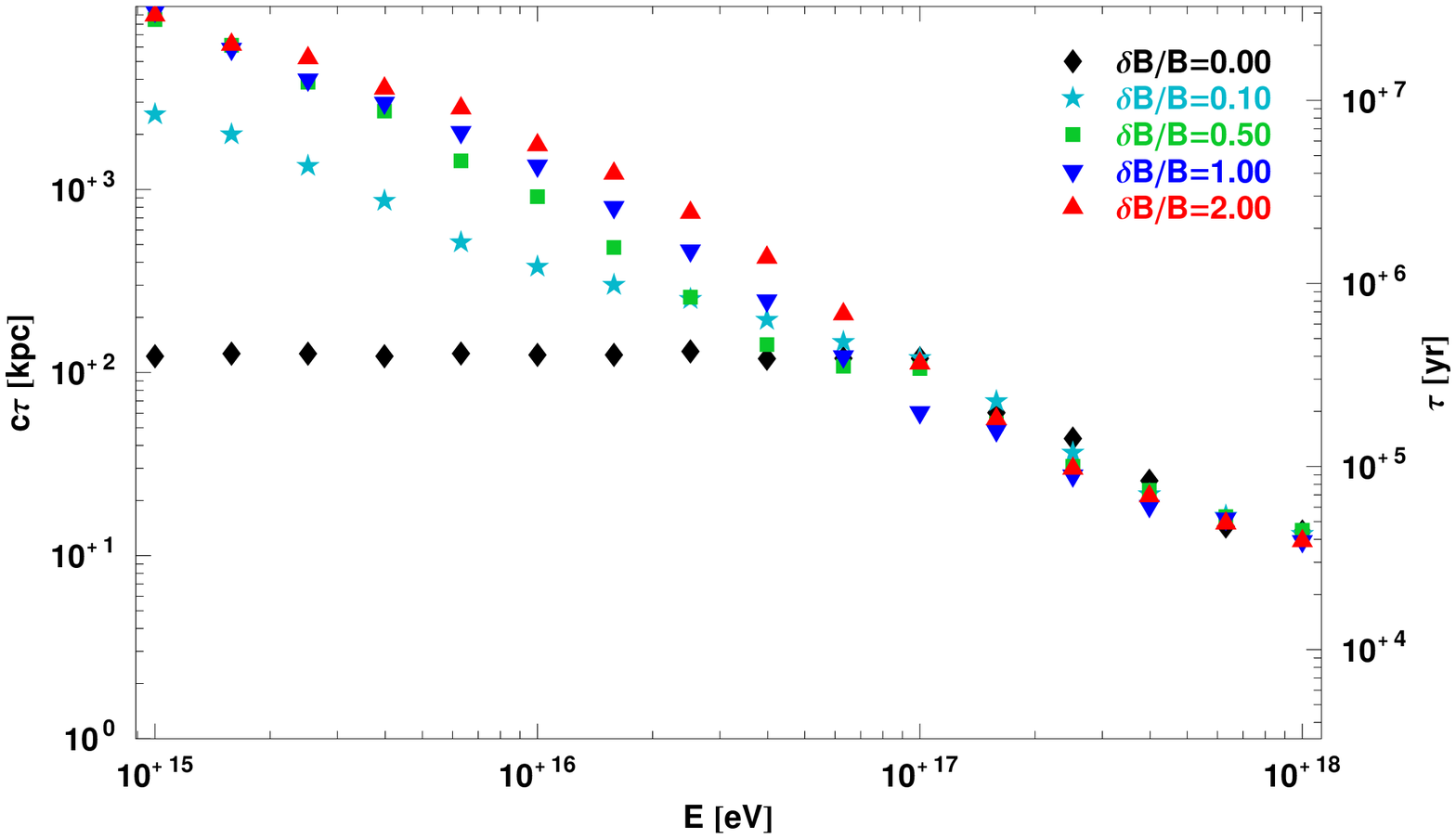}
  \caption{Times of escape in the BSS model for two different ways of normalizing the turbulent
  component of the field. See text for details.}\label{fig:galtoe}
\end{figure}

In Fig.~\ref{fig:galtoe} we plot the residence times of protons injected at the solar system
position and collected at a cylinder with an half-height of $4\kpc$ and a radius of $20\kpc$. In
each panel we plot several series of points corresponding to different levels of turbulence as
indicated. The two panels correspond to two different normalizations of the turbulent component of
the field. In the upper panel the $\sqrt{\langle\delta B^2\rangle}$ of the turbulent field is in
every position a fraction of the regular field. This produces an almost negligible field in the
space between the arms, where the regular field switches direction. In the lower plot the magnitude
of the turbulent field follows the radial and $z$ dependence of the regular field, except for the
arms that are absent in the turbulent component. This means that in this case the ratio $\delta B/B$
is not constant, but variable, being the value indicated in the plot in the center of the arms and
going to infinity in the space between the arms.

The times of escape reflect the increased complexity of the model: they are much less regular than
the ones obtained in the toy model. The normalization is however of the same order of magnitude
with times of escape at $10^{15}\eV$ of about $10^{7}\yr$ and grammages of a few $\g\cm^{-2}$. The
slopes for the times of escape are $\sim 0.7$ and there is no sign of a possible transition to a
flatter slope that would help reconcile these values with the ones measured at low energies.

{\bf Acknowledgments.}
The work of D.D.M. and T.S. is funded in part by NASA APT grant NNG04GK86G. The work of P.B. is
partially funded through grant PRIN-2004.

\end{document}